\documentclass[letterpaper, twocolumn]{IEEEtran}

\usepackage[utf8]{inputenc} 	% UTF-8

\usepackage{amssymb}
\usepackage{bbm} % For indicator function
\usepackage{graphicx}
\usepackage{color,psfrag}
\usepackage{MnSymbol}
\usepackage[normalem]{ulem}

\usepackage{tikz}
\usepackage{siunitx}
\usepackage{cite} % Combining multiple citations
\newcommand{\e}[2]{{\mathbb E}_{#1}\left[ #2 \right]}
\newcommand{\s}[2]{{\frac{1}{{#1}}\sum_n^{#1}} {#2}}

\newcommand{\p}{\mathbb P}
\newcommand{\sub}[1]{_{\text{#1}}}

\newcommand{\opc}{\rho\sub{out}}
\newcommand{\pc}{\rho\sub{cont}}

\newcommand{\preg}{P\sub{cont}}
\newcommand{\xreg}{x\sub{cont}}
\newcommand{\prcvd}{P\sub{rcvd}}
\newcommand{\eprcvd}{\hat{P}\sub{rcvd}}
\newcommand{\yrcvd}{y\sub{rcvd}}
\newcommand{\ptran}{P\sub{tran}}
\newcommand{\xtran}{x\sub{tran}}
\newcommand{\pp}{P\sub{p}}

\newcommand{\yp}{y\sub{p}}
\newcommand{\ys}{y\sub{s}}
\newcommand{\nap}{w\sub{p}}
\newcommand{\nas}{w\sub{s}}
\newcommand{\ite}{\theta\sub{I}}
\newcommand{\rs}{R\sub{s}}
\newcommand{\trs}{R\sub{s}}

\newcommand{\gp}{h\sub{p}}

\newcommand{\epgp}{|\hat{h}\sub{p}|^2}
\newcommand{\epgs}{|\hat{h}\sub{s}|^2}
\newcommand{\gs}{h\sub{s}}
\newcommand{\pgp}{|h\sub{p}|^2}
\newcommand{\egs}{\hat{h}\sub{s}}
\newcommand{\pgs}{|h\sub{s}|^2}

\newcommand{\nps}{\sigma^2}

\newcommand{\fsam}{f\sub{s}}
\newcommand{\ttau}{\tilde{\tau}}

\newcommand{\fprcvd}{F_{\eprcvd}}

\newcommand{\fgs}{F_{\epgs}}

\newcommand{\lp}{\lambda\sub{p}}
\newcommand{\ls}{\lambda\sub{s}}
\newcommand{\Ks}{N\sub{s}}

\DeclareMathOperator*{\maxi}{max}

\DeclareMathOperator*{\ncchi}{\mathcal{X}_1^2}

\newtheorem{theorem}{Theorem}

\newtheorem{lemma}{Lemma}

\newtheorem{coro}{Corollary}

\newtheorem{approxi}{Approximation}
\makeatletter
\if@twocolumn
	\newcommand{\figscale}{0.80 \columnwidth}
	\newcommand{\figscalet}{0.76 \columnwidth}
\else
	\newcommand{\figscale}{0.38 \columnwidth}
	\newcommand{\figscalet}{0.38 \columnwidth}
\fi
\makeatother
\ifCLASSOPTIONcompsoc
\usepackage[caption=false,font=normalsize,labelfont=sf,textfont=sf]{subfig}
\else
\usepackage[caption=false,font=footnotesize]{subfig}
\fi

\usepackage{hyperref}
\allowdisplaybreaks

\begin{document}
\title{Performance Analysis of Underlay Cognitive Radio Systems: Estimation-Throughput Tradeoff}
\author{Ankit Kaushik\IEEEauthorrefmark{1}, \IEEEmembership{Student Member, IEEE},\\ Shree Krishna Sharma\IEEEauthorrefmark{2},  \IEEEmembership{Member, IEEE}, Symeon Chatzinotas\IEEEauthorrefmark{2}, \IEEEmembership{Senior Member, IEEE},  \\ Bj\"orn Ottersten\IEEEauthorrefmark{2}, \IEEEmembership{Fellow, IEEE}, Friedrich K. Jondral\IEEEauthorrefmark{1} \IEEEmembership{Senior Member, IEEE}
\thanks{\IEEEauthorrefmark{1}A. Kaushik and F. K. Jondral are with Communications Engineering Lab, Karlsruhe Institute of Technology (KIT), Germany. Email:{\{ankit.kaushik,friedrich.jondral\}@kit.edu.}}
\thanks{\IEEEauthorrefmark{2}S.K. Sharma, S. Chatzinotas and B. Ottersten are with SnT - securityandtrust.lu, University of Luxembourg, Luxembourg. Email:{\{shree.sharma, symeon.chatzinotas, bjorn.ottersten\}@uni.lu}.}
\thanks{This work was partially supported by the National Research Fund, Luxembourg under the CORE projects ``SeMIGod'' and ``SATSENT''.}
}

% make the title area
\maketitle
\thispagestyle{empty}
\pagestyle{empty}

\begin{abstract}
In this letter, we study the performance of cognitive Underlay Systems (USs) that employ power control mechanism at the Secondary Transmitter (ST). Existing baseline models considered for the performance analysis either assume the knowledge of involved channels at the ST or retrieve this information by means of a feedback channel, however, such situations hardly exist in practice. Motivated by this fact, we propose a novel approach that incorporates the estimation of the involved channels at the ST, in order to characterize the performance of USs under realistic scenarios. Moreover, we apply an outage constraint that captures the impact of imperfect channel knowledge, particularly on the interference power received at the primary receiver. Besides this, we employ a transmit power constraint at the ST to determine an operating regime for the US. Finally, we analyze an interesting tradeoff between the estimation time and the secondary throughput allowing an optimized performance of the US. 
%Our analysis yields a suitable estimation time that renders an optimum performance for the US. 
%\imp{important statement} \\
%\ur{urgent or critical} \\
%\ns{not sure if that is true} \\
%\ws{wrong statement} \\
%\fl{flow of the paper} \\
%\un{unclear statement or argument}
\end{abstract}
%%%%%%%%%%%%%%%%%%%%%%%%%%%%%%%%%%%%%%%%%%%%%%%%%%%%%%%%%%%%%%%%%%%%%%%%%%%%%%%%%%%%%%%%%

\section{Introduction}%%%%%%%%%%%%%%%%%%%%%%%%%%%%%%%%%%%%%%%%%%%%%%%%%%%%%%%%%%%%%%%%%%%%%%%%%%%%%%%%%%%%%%%%%
%\subsection{Background}
Cognitive Radio (CR) communication is considered as one of the viable solutions that addresses the problem of spectrum scarcity of future wireless networks. Secondary access to the licensed spectrum can be broadly categorized into different CR paradigms, namely, interweave, underlay and overlay \cite{Goldsmith09}. Among these, underlay and interweave systems are largely associated with techniques that are present at the physical layer, hence, can be considered feasible for hardware deployment. %Due to its ease of deployment, IS is mostly preferred not only for performing theoretical analysis but for practical implementation as well. 
Particularly, interference tolerance capability exhibited by the Underlay Systems (USs) ensure that they do not cause harmful interference to the primary system while performing shared access to the licensed spectrum. Out of the various underlay techniques, power control is one such mechanism under which USs tend to operate below an Interference Threshold (IT) of the Primary Receiver (PR) \cite{Xing07}.

To employ power control, the knowledge of the interference channel between the ST and the PR is of paramount importance. To this end, performance analysis subject to imperfect channel knowledge has received significant attention \cite{Sharma15, Suraweera10, Kim12, Kaushik15}. According to \cite{Suraweera10, Kim12}, the ST attains the channel knowledge over a feedback channel. Since the feedback channel and the ability to demodulate the ST's signal are non-existent in the current primary systems, the hardware feasibility of this approach becomes challenging. To overcome this issue, a novel strategy was proposed in \cite{Kaushik15}, whereby the ST listens to the control-based transmission from the PR and estimates the received power to retrieve the knowledge of the interference channel. The variation due to imperfect channel knowledge, particularly, in interference power received at the PR was captured by means of a constraint on the probability of confidence \cite{Kaushik15}. 

However, the system model described in \cite{Kaushik15} has certain limitations. Since the USs are sensitive only to those variations that exceed the IT, it is reasonable to implement a power control mechanism subject to an outage constraint. Besides that, the transmit power at the ST should not exceed a certain value. Lastly, analyzing the performance of the secondary system in terms of achievable throughput requires the knowledge of access channel between the ST and Secondary Receiver (SR), however, this knowledge is not available at the ST. In this context, the performance analysis of the US that incorporates channel estimation at the ST subject to outage and transmit power constraints is an interesting research problem. %However, it was revealed that the considered constraint is not suitable for long-term performance characterization of the US \cite{Kaushik15}. %In this letter, we propose an outage constraint on the received power at the PR to apply power control at the ST. %Furthermore, we consider the estimation of the \textit{access} channel between the ST and the Secondary Receiver (SR). 
%\subsection{Motivation}
%Knowledge of the channel transmission is an integral part of an underlay system. Therefore to characterize the true performance of the underlay system, it is important to include channel estimation and imperfection induced due to its estimation in the system model. 
%In order to do so, Kaushik \textit{et. al.} \cite{Kaushik15} introduced the estimation throughput tradeoff (ETT) to capture the estimation error. According to \cite{Kaushik15}, ETT corresponds to the probability of confidence of the received power at the primary receiver (PR) and expected throughput at the secondary receiver (SR). However it was shown that, for the fading channel the probability of confidence didn't sustain the desired constraint. Based on that, it was concluded that probability of confidence constraint was inappropriate for determining the performance at the primary receiver. 

%Moreover, the performance analysis in considered in \cite{Kaushik15} represents a short term interference, that is ST executes the power control to respect the interference constraint for each frame. This demonstrates a flexible approach employed by the US. Apart from this, the regulatory may demand the secondary user to deploy an US based on long term interference. According to it, the ST employs a power control mechanism to sustain an interference constraint for large duration. This duration is much large than a frame duration. In this paper, we investigate the pros and cons of power control mechanism that follows a short and long term interference. 

%\subsection{Contributions}
In this letter, we make the following contributions: 
\begin{itemize}
\item 
We propose a novel model that employs a power control mechanism and incorporates channel estimation of the interacting channels, namely interference channel and access channel at the ST. %In order to acquire channel knowledge, we propose received power estimation at the ST.  
%\item impact of imperfect estimation on the performance of the system. 
\item 
Based on the proposed model, we capture the effect of imperfect channel knowledge by employing an outage constraint on the received power at the PR that restrains the interference encountered by a primary system. Subsequently, we investigate a tradeoff between the estimation time and the achievable secondary throughput. %expression of estimation-throughput tradeoff that analyzes the performance of the US. Using this tradeoff, we determine the maximum achievable throughput for the secondary system. %As a result of the analysis, we propose a power control scheme at the ST subject to outage probability constraint at the PR. 
%\item 
%We finally derive the analytical expressions to perform the short-term and long-term analysis of the CR system. 
\item 
In reference to the transmit power constraint at the ST, we characterize an operating regime for the US. 

\end{itemize}

%%%%%%%%%%%%%%%%%%%%%%%%%%%%%%%%%%%%%%%%%%%%%%%%%%%%%%%%%%%%%%%%%%%%%%%%%%%%%%%%%%%%%%%%%
\section{System Model} \label{sec:sys_mod}
%%%%%%%%%%%%%%%%%%%%%%%%%%%%%%%%%%%%%%%%%%%%%%%%%%%%%%%%%%%%%%%%%%%%%%%%%%%%%%%%%%%%%%%%%
\begin{figure}[!t]
\centering
\includegraphics[width = \figscalet]{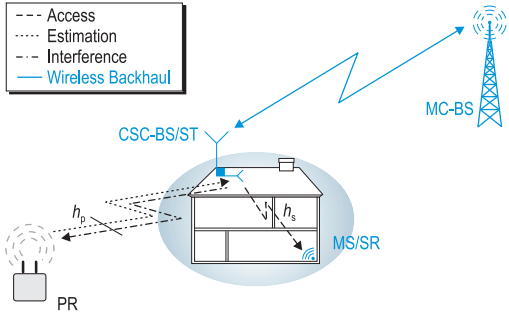}
\caption{A cognitive small cell scenario demonstrating: (i) the underlay paradigm, (ii) the associated network elements, which constitute Cognitive Small Cell-Base Station/Secondary Transmitter (CSC-BS/ST), Mobile Station/Secondary Receiver (MS/SR), Macro Cell-Base Station (MC-BS) and Primary Transmitter (PT), (iii) the interacting channels: interference and access channels.}
\label{fig:scenario}
%\vspace{-5mm}
\end{figure}
%\subsubsection{Medium access}

%\subsection{Underlay scenario}

Cognitive Small Cell (CSC), a CR application, characterizes a small cell deployment that fulfills the spectral requirements for Mobile Stations (MSs) operating indoor, cf. \figurename~\ref{fig:scenario}.
For the disposition of the CSC in the network, the following key elements are essential: a CSC-Base Station (CSC-BS), a Macro Cell-Base Station (MC-BS) and MS \cite{Kaushik16}.
Considering the fact that the power control is employed at the CSC-BS, the CSC-BS and the MS represents ST and SR, respectively. The PR performs transmission/reception of signals (interchangeably over time) to/from the Primary Transmitter (PT), cf. \figurename~\ref{fig:scenario}, the ST follows this alignment for transmitting signals with controlled power over the access channel. %In addition, considering the deployment scenario, due to the outside walls, the SR encounters a loss of $\SI{20}{dB}$ in the received power, 
%Cognitive Relay (CR) \cite{Kaushik14} characterizes a small cell deployment that fulfills the spectral requirements for Indoor Devices (IDs). \figurename~\ref{fig:scenario} illustrates a snapshot of a CR scenario to depict the interaction between the CR with PR and ID, where CR and ID represents the ST and SR respectively. In \cite{Kaushik14}, the challenges involved while deploying the CR as US were presented. However for simplification, a constant transmit power was considered at the CR. Now, we extend the analysis to employ power control at the ST. 
\begin{figure}[!t]
\centering
\includegraphics[width = \figscalet]{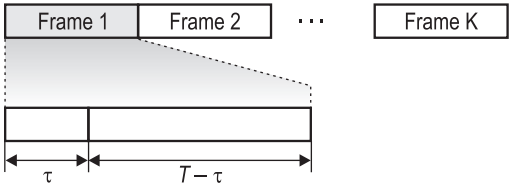}
\caption{Frame structure of underlay system with received power estimation for the interference channel.}
\label{fig:fs}
\vspace{-5mm}
\end{figure}

The medium access for the US is slotted, where the time axis is segmented into frames of length $T$. The duration $T$ ($\ll$ expected ON/OFF period of the primary user) is chosen in such a way that the frames are aligned to the control-based transmissions\footnote{These include the beacons transmitted by the PR in the same band or a pilot channel transmitted in a separate band. With the knowledge of power transmitted by the PR and using channel reciprocity, the ST is able to determine the interference power received at the PR, thus, can control its transmit power while sustaining the IT, cf. \cite{Kaushik15} and references therein.} of the primary system. The frame structure is analog to periodic sensing in \cite{Kaushik15}, according to which, US uses a time interval $\tau (< T)$ to estimate the received power, cf. \figurename~\ref{fig:fs}. In order to consider variations due to channel fading, we assume that the interacting channels remain constant over a frame duration \cite{Kaushik15}. Hence characterized by the fading process, each frame witnesses a different received power. 

In order to operate at a desired outage probability, defined later in Section \ref{sec:et_ana}, it is reasonable to exercise estimation followed by data transmission with power control in the remaining time $T - \tau$ for each frame. %At the end, the remaining time $T - \tau$ is utilized for data transmission with controlled power. 
Since the access channel estimation is performed by listening to the pilot symbols from the SR, no time resources are allocated for access channel estimation in the frame structure, hence $\tau$ is utilized for interference channel estimation only. 
In this letter, we consider this frame structure to perform short-term analysis, i.e., the performance is analyzed for a certain channel gain, without taking into account the effect of channel fading.%, and long-term analysis, which includes channel fading. Depending on their applicability, the aforementioned analyses can be considered for the hardware deployment. 

To simplify the analysis, we assume that during data transmission at the ST, the interference at the SR, from the PT, to be below the noise level. However, by replacing the noise power with interference plus noise power in the throughput expressions, derived later in Section \ref{sec:et_ana}, the performance of the US under the interference limited regime can be depicted. %In this view, the interaction with the PT is excluded in the considered scenario, cf. \figurename~\ref{fig:scenario}. 

%\subsection{Signal model}
In the estimation phase, the discrete control-based signal received from the PR at the ST is given by \cite{Kaushik15}
\begin{equation}
\yrcvd[n] = \gp \cdot \xtran[n] + \nas[n],
\label{eq:sys_mod_st}
\end{equation}
where $\xtran[n]$ corresponds to a discrete and complex sample transmitted by the PR with transmit power $\ptran$ known at the ST, $\pgp$ represents the power gain for the interference channel and $\nas[n]$ is circularly symmetric complex Additive White Gaussian Noise (AWGN) at the ST with %The transmitted power at PR is $\ptran$, considering that $\tau \fsam$ $(= N)$ is the number of samples used for estimation. 
$\mathcal{CN}(0, \nps)$. %The power received at the ST is given as 
%\begin{align}
%\prcvd = \s{\tau \fsam}{ |\sqrt{\gp \ap} \xtran[n] + \nas[n]|^2}.
%\label{eq:prcvd} 
%\end{align}

During data transmission phase, the interference signal received at the PR is given by
\begin{equation}
\yp[n] = \gp  \cdot \xreg[n] + \nap[n],
\label{eq:sys_mod_pr}
\end{equation}
and on the other side, the received signal at the SR follows 
\begin{equation}
\ys[n] = \gs \cdot \xreg[n] + \nas[n],
\label{eq:sys_mod_sr}
\end{equation}
where $\xreg[n]$ corresponds to a discrete and complex sample transmitted by the ST with controlled power $\preg$. Further, $\pgs$ represents the power gain for the access channel and $\nap[n]$ is AWGN at the PR with $\mathcal{CN}(0, \nps)$. %The power received at the ST is given as 
%Considering (\ref{eq:sys_mod_pr}), the powers received at the PR and SR are evaluated as $\pp = \s{(T - \tau) \fsam}{|\yp[n]^2|}$ and $\ps = \s{(T - \tau) \fsam}{|\yp[n]^2|}$. Likewise (\ref{eq:sys_mod_st}), $\nap[n]$ and $\nas[n]$ represents circularly symmetric AWGN at PR and ST with zero mean and variance $\e{}{|\nap[n]|^2} = \npp$ and $\e{}{|\nas[n]|^2} = \nps$ correspondingly. Consider that $\ptran$, $\preg$ and $\pp$ correspond to power for a given frame. 
%\subsubsection{Channel}

%We consider that all transmitted signals are subjected to distance dependent path loss. The small scale fading gains $\gp, \gs$ are modelled as frequency-flat fading. Hence, the $\gp, \gs$ follow an exponential distribution \cite{Tse05} where $\e{}{\gp}$ and $\e{}{\gs}$ represent their path loss.

%%%%%%%%%%%%%%%%%%%%%%%%%%%%%%%%%%%%%%%%%%%%%%%%%%%%%%%%%%%%%%%%%%%%%%%%%%%%%%%%%%%%%%%%%
\section{Performance Analysis} \label{sec:et_ana}
%%%%%%%%%%%%%%%%%%%%%%%%%%%%%%%%%%%%%%%%%%%%%%%%%%%%%%%%%%%%%%%%%%%%%%%%%%%%%%%%%%%%%%%%%
%\subsection{Short-term Analysis}
%We first investigate the short-term analysis, whereby the channel is $\gp, \gs$ to a and unknown.
\subsection{Ideal model}
According to the ideal model, a ST as an US is required to control its transmit power in such a way that the interference power received $\pp$ at the PR is below IT $\ite$ \cite{Xing07}
\begin{equation}
\pp = \pgp \preg \le \ite.
\label{eq:IT_id}
\end{equation}
%where $\alpha$ denotes the distance dependent path loss. $g\sub{p}$ represents the small-scale channel fading. 

With controlled power at the ST determined using (\ref{eq:IT_id}), the throughput at the SR is defined as
\begin{equation}
\rs = \log_2 \left(1 + \pgs \frac{\preg}{ \nps} \right). 
\label{eq:Thr_id}
\end{equation}
%where $\e{\gs, \gp}{\cdot}$ represents the expectation over $\gs, \gp$. 
\subsection{Proposed Model} 
To employ power control based on (\ref{eq:IT_id}) and evaluate $\rs$ according to (\ref{eq:Thr_id}), the ideal model considers the knowledge of the involved channels $\gp$ and $\gs$ at the ST, which is not available in practice. In this regard, we incorporate channel estimation in the system model. The imperfect channel knowledge, however, translates to variations in the performance parameters, $\pp$ and $\trs$. Particularly, a variation in $\pp$ that exceeds the $\ite$ causes interference at the PR. Unless characterized, these variations may seriously degrade the performance of the US. In this view, we capture the variations in $\pp$ and $\trs$ by characterizing the distribution functions of the estimated channels.

\subsubsection{Estimation of interference channel}
Given $\prcvd = \pgp \ptran + \nps$ and the knowledge of PR's transmit power $\ptran$, the ST listens to the control-based transmissions from the PR and acquires the knowledge of $\epgp$ indirectly by estimating the received power $\eprcvd = \s{\tau \fsam}{ |\yrcvd[n]|^2}$.
%\begin{align}
%\eprcvd = \s{\tau \fsam}{ |\yrcvd[n]|^2}.
%\label{eq:prcvd} 
%\end{align}
$\eprcvd$ estimated using $\tau \fsam$ samples follows a non-central chi-squared distribution $\fprcvd \sim \ncchi(\lp, \tau \fsam)$ with non-centrality parameter $\lp = \tau \fsam \pgp \ptran /\nps = \tau \fsam \gamma$, where $\gamma$ is defined as the ratio of the received control-based power (from the PR) to noise at the ST and $\tau \fsam$ corresponds to the degrees of freedom.
\begin{approxi} \label{ap:ap1}
For all degrees of freedom, the $\ncchi$ distribution can be approximated by a Gamma distribution \cite{abramo}. The parameters of the Gamma distribution are obtained by matching the first two central moments to those of $\ncchi$.
\end{approxi}
\begin{lemma} \label{lm:lm1}
The cumulative distribution function of $\eprcvd$ is characterized as 
\begin{align}
\fprcvd(x) \approx 1 - \Gamma(&a_1, b_1 x) \label{eq:fprcvd},  
\\ \text{where  } a_1 = \frac{\tau \fsam (1 + \gamma)^2}{2 + 4 \gamma} &\text{ and } b_1 = \frac{\nps (2 + 4 \gamma)}{\tau \fsam (1 + \gamma)},  \nonumber 
\end{align} 
%$\text{where  } a_1 = \frac{\tau \fsam (1 + \gamma)^2}{2 + 4 \gamma} \text{ and } b_1 = \frac{\nps (2 + 4 \gamma)}{\tau \fsam (1 + \gamma)}$   
and $\Gamma(\cdot, \cdot)$ represents the regularized lower-incomplete Gamma function \cite{abramo}. 
\end{lemma}
\begin{IEEEproof}
Applying Approximation \ref{ap:ap1} to $\ncchi(\lp, \tau \fsam)$ yields (\ref{eq:fprcvd}). 
\end{IEEEproof}
\subsubsection{Estimation of access channel}
The pilot signal received from the SR undergoes matched filtering and demodulation at the ST, hence, we employ a pilot-based estimation at the ST to acquire the knowledge of the access channel. According to \cite{Gifford08}, the maximum-likelihood estimate with $\Ks$ pilot symbols is given by 
\begin{align}
\gs = \egs + \frac{\sum^{\Ks}_{n} p[n]}{2 \Ks},
\label{eq:pilot_MLE}
\end{align}
where $p[n]$ denotes the discrete pilot symbol and $\frac{\sum^{\Ks}_{n} p[n]}{2 \Ks}$ represents the estimation error.
%(\ref{eq:pilot_MLE}) illustrates a correlation between the $\hs$ and $\ehs$. 
As a result, the estimate $\egs$ is unbiased, efficient, i.e., achieves the Cram\'er-Rao bound with equality, with asymptotic variance $\e{}{|\gs -\egs|^2} = \frac{\nps}{2 \Ks}$ \cite{Gifford08}. Hence, $\egs$ conditioned on $\gs$ follows a Gaussian distribution
\begin{align}
\egs|\gs \sim \mathcal{N}\left( \gs,\frac{\nps}{2 \Ks} \right).
\label{eq:ehs} 
\end{align}
Consequently, the estimated power gain $|\egs|^2$ follows a non-central chi-squared $\ncchi(\ls, 1)$ distribution with 1 degree of freedom and non-centrality parameter $\ls = \frac{2 \Ks |\gs|^2}{\nps}$. 
\begin{lemma} \label{lm:lm2}
The cumulative distribution function of $\epgs$ is characterized as 
\begin{align}
\fgs(x) \approx 1 - \Gamma(&a_2, b_2 x) \label{eq:fehs},\\ 
\text{where  } a_2 = \frac{(1 + \ls)^2}{2 + 4 \ls} &\text{ and } b_2 = \frac{\nps (2 + 4 \ls)}{(1 + \ls)}.  \nonumber 
\end{align} 
\end{lemma}
\begin{IEEEproof}
Applying Approximation \ref{ap:ap1} to $\ncchi(\ls,1 )$ yields (\ref{eq:fehs}). 
\end{IEEEproof}
%  desired by the regulatory bodies, we capture this variation by means of an outage probability constraint $\p(\pp \ge \ite) \le \opc$.  %, it is important for the system to restrain $\op$ above a certain desired level $\opc$. %As a result, the outage probability constraint is defined as
%\begin{align}
%\p(\pp = \gp \preg \ge \ite) \le \opd. 
%\label{eq:opc} 
%\end{align} 
%\begin{lemma}
%The controlled power that satisfies 
%\end{lemma}
%\subsubsection{Power control}
%Given $\tau$ is utilized for $\prcvd$ estimation, the throughput at the SR is given by  
%\begin{equation}
%\rs(\tau) = \frac{T - \tau}{T} \log_2 \left(1 + \frac{\gs \preg(\tau) }{\nps} \right). 
%\label{eq:Thr_pm}
%\end{equation}
%It will be clear later in this section that small $\tau$ results in large variations for the $\prcvd$ and vice versa. According to (\ref{eq:preg}), this induces variations in $\preg$. These variations causes $\pp$ to deviate from $\ite$. If not considered in the model, these variations may affect the performance of the system. 
%Based on (\ref{eq:opc}), the controlled power at the PR is determined as 
%\begin{align}
%\preg(\tau) = \ite \cdot F_{\frac{1}{\gp}}^{-1}(\opd, \tau), 
%\label{eq:preg}
%\end{align}
%where $\mathcal{F}_{\frac{1}{\gp}}^{-1}$ represents the inverse-distribution function of $1/{\gp}$.
%Clearly, there exits a tradeoff that involves maximizing the expected throughput at the SR subject to a outage probability constraint is given by 
Next, we employ an outage probability constraint at the ST to capture the variation in the $\pp$ incurred due to channel estimation, defined as 
\begin{align}
\p\left( \left( \smash[b]{\underbrace{\frac{\eprcvd - \nps}{\ptran}}_{\epgp}} \right) \preg \ge \ite \right) \le \opc, \label{eq:opc} \\[0.1em] \nonumber 
\end{align}
where $\opc$ corresponds to an outage constraint. 
Besides the outage constraint, $\preg$ is limited by a predefined transmit power $\pc$. To capture this aspect, the transmit power constraint at the ST is defined as
\begin{align}
\preg \le \pc. \label{eq:pc} 
\end{align} 
Based on the aforementioned constraints, we determine the expression of controlled power for the proposed model.
\begin{lemma} \label{lm:lm3}
Subject to the outage constraint and transmit power constraint, the controlled power at the ST is given by
\begin{align}
\preg &= 
\begin{cases} 
\frac{\ite \ptran}{ \left(b_1 \Gamma^{-1}(1-\opc, a_1) - \nps  \right)} & \mbox{if } \preg < \pc \\
\pc & \mbox{if } \preg \ge \pc
\end{cases},
\label{eq:preg} 
\end{align}
where $a_1$ and $b_1$ are defined in (\ref{eq:fprcvd}) and $\Gamma^{-1}(\cdot, \cdot)$ is the inverse function of regularized lower-incomplete Gamma function \cite{abramo}.
\end{lemma} 
\begin{IEEEproof}
Substituting the distribution function for $\eprcvd$, defined in (\ref{eq:fprcvd}), in (\ref{eq:opc}) and combining with (\ref{eq:pc}) yields (\ref{eq:preg}).
\end{IEEEproof}

\begin{figure}[!t]

%% Add psfrag entries
% This file is generated by the MATLAB m-file laprint.m. It can be included
% into LaTeX documents using the packages graphicx, color and psfrag.
% It is accompanied by a postscript file. A sample LaTeX file is:
%    \documentclass{article}\usepackage{graphicx,color,psfrag}
%    \begin{document}\input{fig_N_vs_SNR_diff_pout_diff_maxContPow_th}\end{document}
% See http://www.mathworks.de/matlabcentral/fileexchange/loadFile.do?objectId=4638
% for recent versions of laprint.m.
%
% created by:           LaPrint version 3.16 (13.9.2004)
% created on:           31-Aug-2015 16:25:39
% eps bounding box:     16 cm x 12 cm
% comment:              
%
%\begin{psfrags}%
%\psfragscanon%
%
% text strings:
\psfrag{s05}[b][b]{\fontsize{8.5}{12.75}\fontseries{m}\mathversion{normal}\fontshape{n}\selectfont \color[rgb]{0,0,0}\setlength{\tabcolsep}{0pt}\begin{tabular}{c}$\tau$ [ms]\end{tabular}}%
\psfrag{s06}[t][t]{\fontsize{8.5}{12.75}\fontseries{m}\mathversion{normal}\fontshape{n}\selectfont \color[rgb]{0,0,0}\setlength{\tabcolsep}{0pt}\begin{tabular}{c}$\gamma$ [dB]\end{tabular}}%
\psfrag{s10}[][]{\fontsize{10}{15}\fontseries{m}\mathversion{normal}\fontshape{n}\selectfont \color[rgb]{0,0,0}\setlength{\tabcolsep}{0pt}\begin{tabular}{c} \end{tabular}}%
\psfrag{s11}[][]{\fontsize{10}{15}\fontseries{m}\mathversion{normal}\fontshape{n}\selectfont \color[rgb]{0,0,0}\setlength{\tabcolsep}{0pt}\begin{tabular}{c} \end{tabular}}%
\psfrag{s12}[l][l]{\fontsize{8.5}{12.75}\fontseries{m}\mathversion{normal}\fontshape{n}\selectfont \color[rgb]{0,0,0}$\opc = 0.1$}%
\psfrag{s13}[l][l]{\fontsize{8.5}{12.75}\fontseries{m}\mathversion{normal}\fontshape{n}\selectfont \color[rgb]{0,0,0}$\opc = 0.01$}%
\psfrag{s14}[l][l]{\fontsize{8.5}{12.75}\fontseries{m}\mathversion{normal}\fontshape{n}\selectfont \color[rgb]{0,0,0}$\opc = 0.1$}%
%
% axes font properties:
\fontsize{8.5}{12.75}\fontseries{m}\mathversion{normal}%
\fontshape{n}\selectfont%
%
% xticklabels:
\psfrag{x01}[t][t]{-20}%
\psfrag{x02}[t][t]{-18}%
\psfrag{x03}[t][t]{-16}%
\psfrag{x04}[t][t]{-14}%
\psfrag{x05}[t][t]{-12}%
\psfrag{x06}[t][t]{-10}%
\psfrag{x07}[t][t]{-8}%
\psfrag{x08}[t][t]{-6}%
\psfrag{x09}[t][t]{-4}%
\psfrag{x10}[t][t]{-2}%
\psfrag{x11}[t][t]{0}%
\psfrag{x12}[t][t]{2}%
%
% yticklabels:
\psfrag{v01}[r][r]{0}%
\psfrag{v02}[r][r]{2}%
\psfrag{v03}[r][r]{4}%
\psfrag{v04}[r][r]{6}%
\psfrag{v05}[r][r]{8}%
\psfrag{v06}[r][r]{10}%
\psfrag{v07}[r][r]{12}%
\psfrag{v08}[r][r]{14}%
\psfrag{v09}[r][r]{16}%
\psfrag{v10}[r][r]{18}%
\psfrag{v11}[r][r]{20}%
%
% Figure:
%\resizebox{8cm}{!}{\includegraphics{fig_N_vs_SNR_diff_pout_diff_maxContPow_th.eps}}%
%\end{psfrags}%
%
% End fig_N_vs_SNR_diff_pout_diff_maxContPow_th.tex

\centering
\begin{tikzpicture}[scale=1]
\node[anchor=south west,inner sep=0] (image) at (0,0)
{
\includegraphics[width= \figscale]{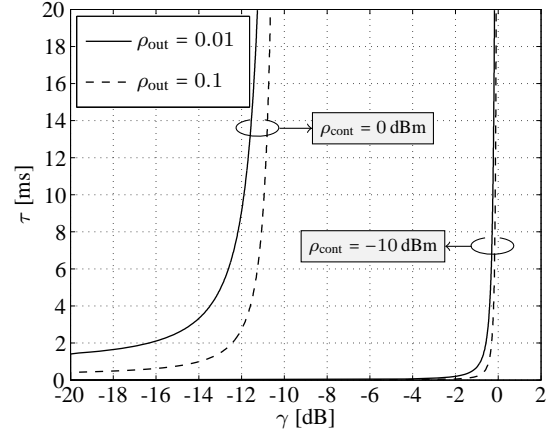}
};
\begin{scope}[x={(image.south east)},y={(image.north west)}]

\draw (0.86,0.42) arc(-250:70:0.04 and 0.02);
\node[draw,fill=gray!10,font=\scriptsize] (text1) at (0.65,0.4) {$\pc = \SI{-10}{dBm}$};
\draw[black, <-] (text1.east) -- (0.835,0.4);
\draw (0.42,0.7) arc(-250:70:0.04 and 0.02);
\node[draw,fill=gray!10,font=\scriptsize] (text2) at (0.65,0.68) {$\pc = \SI{0}{dBm}$};
\draw[black, <-] (text2.west) -- (0.475,0.68);

%\draw[help lines,xstep=.1,ystep=.1] (0,0) grid (1,1);
%\foreach \x in {0,1,...,9} { \node [anchor=north] at (\x/10,0) {0.\x}; }
%\foreach \y in {0,1,...,9} { \node [anchor=east] at (0,\y/10) {0.\y}; }
\end{scope}
\end{tikzpicture}

\caption{An illustration of operating regime ($\gamma^*$) for the US depicted in terms of estimation time ($\tau$) and the ratio of the received control-based power (from the PR) to noise ($\gamma$) at the ST.}
\label{fig:or}
\vspace{-5mm}
\end{figure}
Clearly, $\preg$ increases with increase in $\pgs$, which depicts low $\gamma$, consequently a better performance in terms of secondary throughput is achieved by the US for low $\gamma$, however with the presence to $\pc$ an upper limit is imposed on the achievable performance. We define this performance limit in terms of $\gamma$ as an operating regime $\gamma^*$ for the US. 
\begin{coro}
Subject to a transmit power constraint, an operating regime at the ST is defined as\footnote{Please note that $\tau \fsam$ and $\gamma$ are included in the parameters $a_1$ and $b_1$, cf. (\ref{eq:fprcvd}).} 
\begin{align}
\opc \le 1 - \Gamma\left(a_1, \frac{1}{b_1} \left( \frac{\ite \ptran}{\pc} + \nps  \right)  \right). \label{eq:opreg}  
\end{align}
\end{coro}
\begin{IEEEproof}
Substituting $\preg$, cf. (\ref{eq:preg}), in (\ref{eq:pc}) results in (\ref{eq:opreg}). Replacing $(\ref{eq:opreg})$ with equality yields $\gamma^*$. 
\end{IEEEproof}
In other words, below a certain $\gamma (\le \gamma^{*})$ no performance gain is witnessed by the CR system, cf. \figurename~\ref{fig:or}. As a result, by replacing $\gamma^*$ in the following expression of secondary throughput, we determine the performance limits of operation for the US. 
%Based on the operation regime, we determine best performance, in terms of the lowest operational SNR, achieved by the US. 
 
Besides that, for the estimation model, the expected throughput for the access link at the SR is defined as
\begin{align}
\rs(\tau) = \e{\epgs} {\frac{T - \tau}{T} \log_2 \left(1 + \frac{\epgs \preg }{\nps} \right)}, \label{eq:rs}
\end{align} 
where $\e{\epgs}{\cdot}$ corresponds to an expectation over $\epgs$, whose distribution function is characterized in Lemma \ref{lm:lm2}. 

At this stage, it is worthy to note that $\preg$ and $\trs$ depend on $\tau$, cf. (\ref{eq:preg}) and (\ref{eq:rs}), respectively. Hence, the proposed model exhibits a fundamental tradeoff between the estimation time and the achievable secondary throughput.  
\begin{theorem} \label{th:th1}
\normalfont
The expected achievable secondary throughput subject to the outage constraint on the received power at the PR and transmit power constraint at the ST is defined as
\begin{align}
\trs(\ttau) = \maxi_{\tau}  & \text{      } {\rs(\tau)}, 
 \label{eq:sys} \\
\text{s.t.} & \text{ } (\ref{eq:opc}), \text{  } (\ref{eq:pc}), \nonumber 
%\text{s.t.} & \text{ } \p\left( \left( \frac{\eprcvd - \nps}{\ptran} \right) \preg \ge \ite \right) \le \opc, \nonumber \\ 
%\text{s.t.} & \text{ }  \preg \le \pc, \nonumber   
 \end{align}
where $\trs(\ttau)$ corresponds to optimum throughput at $\ttau$.  
\end{theorem}
\begin{IEEEproof}
The constrained optimization problem is solved by substituting $\preg$ from Lemma \ref{lm:lm3}, determined by applying outage and transmit power constraints defined in (\ref{eq:opc}) and (\ref{eq:pc}), in (\ref{eq:rs}). 

Using the distribution function of $\epgs$ in (\ref{eq:fehs}) to determine an expression of expected throughput as a function of $\tau$. Solving numerically this expression yields $\ttau$ and $\trs(\ttau)$. 
\end{IEEEproof}
%Theorem \ref{th:th1} depicts an optimum estimation time $\ttau$ that achieves an optimum secondary throughput $\trs(\ttau)$.
%The throughput depicted from (\ref{eq:Thr_id}) for the ideal model overestimates the throughput of the US. %This overestimation in the throughput is evaluated as $\beta$ that depicts the difference between the $\ers$ obtained from the models. 
%It is evident that for analyzing the tradeoff depicted in (\ref{eq:sys}), first it is important to characterize the $\preg$ and density function $\drs$. 

%\subsection{Short term analysis}
%To simplify the analysis, we consider a case where the transmitted signals are subject to path loss only, that is, the small scale channel gains correspond to $\gs = \gp = 1$. In this way, we first consider the variations in $\pp$ due to the presence of noise in the system. 

%\subsubsection{Characterization of performance parameters}
%Considering (\ref{eq:prcvd}), $\prcvd$ follows a non-central chi-squared distributed $\mathcal{X'}^2(N \gamma, N)$, where $\gamma = \ptran/\nps$ denotes the received SNR \cite{Urkowitz}. 
%According to (\ref{eq:preg}), $\preg$ for the short term case is given by 

%\subsection{Long-term analysis}
 
%%%%%%%%%%%%%%%%%%%%%%%%%%%%%%%%%%%%%%%%%%%%%%%%%%%%%%%%%%%%%%%%%%%%%%%%%%%%%%%%%%%%%%%%%
\section{Numerical Analysis} \label{sec:num_ana}
%%%%%%%%%%%%%%%%%%%%%%%%%%%%%%%%%%%%%%%%%%%%%%%%%%%%%%%%%%%%%%%%%%%%%%%%%%%%%%%%%%%%%%%%%
Here, we investigate the performance of the US based on the proposed model. To accomplish this: (i) we perform simulations to validate the expressions obtained, (ii) we analyze the performance loss incurred due to the estimation. In this regard, we consider the ideal model for benchmarking and evaluating the performance loss. % (iii) we establish mathematical justification to the considered approximations.
% Unless stated explicitly, the following choice of the parameters is considered for the analysis, cf. Table \ref{tb:tb2}.
Unless stated explicitly, the following choice of the parameters is considered for the analysis, $\fsam = \SI{1}{MHz}$, $\gp = \SI{-100}{dB}$, $\gs = \SI{-80}{dB}$, $\ite = \SI{-110}{dBm}$, $T = \SI{100}{ms}$, $\opc \in \{0.01, 0.10\}$, $\pc \in \{-10, 0\} \SI{}{dBm}$, $\nps = \SI{-100}{dBm}$, $\gamma = \SI{0}{dB}$, $\ptran = \SI{0}{dBm}$, $\Ks = 10$.

%\subsection{Short-term analysis}

%\begin{table}
%%\vspace{-0.4cm}
%\renewcommand{\arraystretch}{1.2}
%\caption{Parameters for Numerical Analysis}
%%\vspace{-0.6cm}
%\label{tb:tb2}
%\centering
%\scriptsize{
%\begin{tabular}{c||c}
%\hline
%\bfseries Parameter & \bfseries Value \\
%\hline\hline
%$\fsam$  & $\SI{1}{MHz}$ \\ 
%$\gp$ & $\SI{-100}{dB}$ \\ 
%$\gs$ & $\SI{-80}{dB}$ \\ 
%$\ite$ & $\SI{-110}{dB}$ \\ 
%$T$ & $\SI{100}{ms}$ \\ 
%$\opc$ & \{0.01, 0.10\} \\ 
%$\pc$ & \SI{0}{dBm} \\ 
%$\nps$ & $\SI{-100}{dBm}$ \\ 
%$\gamma$ & $\SI{0}{dB}$ \\ 
%$\ptran$ & $\SI{0}{dBm}$ \\ 
%$\Ks$ & 10 \\ \hline
%\end{tabular}}
%\end{table}

\begin{figure}[!t]
%% Add psfrag entries
% This file is generated by the MATLAB m-file laprint.m. It can be included
% into LaTeX documents using the packages graphicx, color and psfrag.
% It is accompanied by a postscript file. A sample LaTeX file is:
%    \documentclass{article}\usepackage{graphicx,color,psfrag}
%    \begin{document}\input{fig_thr_est_time_tradeoff_AWGN}\end{document}
% See http://www.mathworks.de/matlabcentral/fileexchange/loadFile.do?objectId=4638
% for recent versions of laprint.m.
%
% created by:           LaPrint version 3.16 (13.9.2004)
% created on:           15-Sep-2015 11:55:33
% eps bounding box:     16 cm x 12 cm
% comment:              
%
%\begin{psfrags}%
%\psfragscanon%
%
% text strings:
\psfrag{s05}[b][b]{\fontsize{8.5}{12.75}\fontseries{m}\mathversion{normal}\fontshape{n}\selectfont \color[rgb]{0,0,0}\setlength{\tabcolsep}{0pt}\begin{tabular}{c}$\rs(\tau)$ [bits/sec/Hz]\end{tabular}}%
\psfrag{s06}[t][t]{\fontsize{8.5}{12.75}\fontseries{m}\mathversion{normal}\fontshape{n}\selectfont \color[rgb]{0,0,0}\setlength{\tabcolsep}{0pt}\begin{tabular}{c}$\tau$ [ms]\end{tabular}}%
\psfrag{s10}[][]{\fontsize{10}{15}\fontseries{m}\mathversion{normal}\fontshape{n}\selectfont \color[rgb]{0,0,0}\setlength{\tabcolsep}{0pt}\begin{tabular}{c} \end{tabular}}%
\psfrag{s11}[][]{\fontsize{10}{15}\fontseries{m}\mathversion{normal}\fontshape{n}\selectfont \color[rgb]{0,0,0}\setlength{\tabcolsep}{0pt}\begin{tabular}{c} \end{tabular}}%
\psfrag{s12}[l][l]{\fontsize{8.5}{12.75}\fontseries{m}\mathversion{normal}\fontshape{n}\selectfont \color[rgb]{0,0,0}Simulated}%
\psfrag{s13}[l][l]{\fontsize{8.5}{12.75}\fontseries{m}\mathversion{normal}\fontshape{n}\selectfont \color[rgb]{0,0,0}IM}%
\psfrag{s14}[l][l]{\fontsize{8.5}{12.75}\fontseries{m}\mathversion{normal}\fontshape{n}\selectfont \color[rgb]{0,0,0}EM, $\opc = 0.01$}%
\psfrag{s15}[l][l]{\fontsize{8.5}{12.75}\fontseries{m}\mathversion{normal}\fontshape{n}\selectfont \color[rgb]{0,0,0}EM, $\opc = 0.1$}%
\psfrag{s16}[l][l]{\fontsize{8.5}{12.75}\fontseries{m}\mathversion{normal}\fontshape{n}\selectfont \color[rgb]{0,0,0}$\trs(\ttau)$}%
\psfrag{s17}[l][l]{\fontsize{8.5}{12.75}\fontseries{m}\mathversion{normal}\fontshape{n}\selectfont \color[rgb]{0,0,0}Simulated}%
%
% axes font properties:
\fontsize{8.5}{12.75}\fontseries{m}\mathversion{normal}%
\fontshape{n}\selectfont%
%
% xticklabels:
\psfrag{x01}[t][t]{0}%
\psfrag{x02}[t][t]{1}%
\psfrag{x03}[t][t]{2}%
\psfrag{x04}[t][t]{3}%
\psfrag{x05}[t][t]{4}%
\psfrag{x06}[t][t]{5}%
\psfrag{x07}[t][t]{6}%
\psfrag{x08}[t][t]{7}%
\psfrag{x09}[t][t]{8}%
\psfrag{x10}[t][t]{9}%
\psfrag{x11}[t][t]{10}%
%
% yticklabels:
\psfrag{v01}[r][r]{2.4}%
\psfrag{v02}[r][r]{2.5}%
\psfrag{v03}[r][r]{2.6}%
\psfrag{v04}[r][r]{2.7}%
\psfrag{v05}[r][r]{2.8}%
\psfrag{v06}[r][r]{2.9}%
\psfrag{v07}[r][r]{3}%
\psfrag{v08}[r][r]{3.1}%
\psfrag{v09}[r][r]{3.2}%
\psfrag{v10}[r][r]{3.3}%
\psfrag{v11}[r][r]{3.4}%
%
% Figure:
%\resizebox{8cm}{!}{\includegraphics{fig_thr_est_time_tradeoff_AWGN.eps}}%
%\end{psfrags}%
%
% End fig_thr_est_time_tradeoff_AWGN.tex

\centering
\includegraphics[width= \figscale]{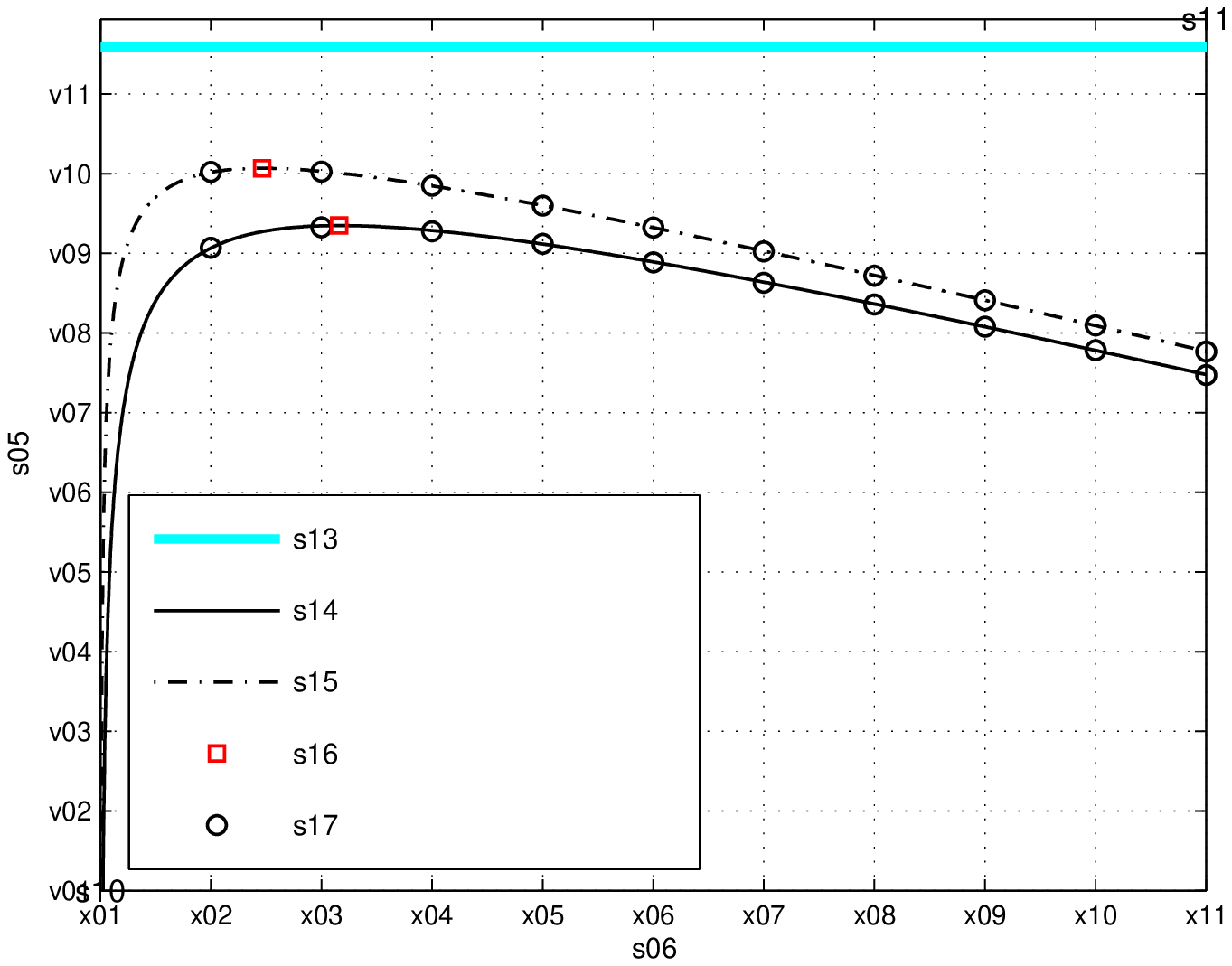}
\caption{Estimation-throughput tradeoff with $\gamma = \SI{0}{dB}$, $\opc \in \{0.01, 0.1\}$ and $\pc = \SI{0}{dBm}$.}
\label{fig:ETT}
\vspace{-7mm}
\end{figure}

\begin{figure}[!t]
%% Add psfrag entries
% This file is generated by the MATLAB m-file laprint.m. It can be included
% into LaTeX documents using the packages graphicx, color and psfrag.
% It is accompanied by a postscript file. A sample LaTeX file is:
%    \documentclass{article}\usepackage{graphicx,color,psfrag}
%    \begin{document}\input{fig_opt_thr_vs_SNR_AWGN}\end{document}
% See http://www.mathworks.de/matlabcentral/fileexchange/loadFile.do?objectId=4638
% for recent versions of laprint.m.
%
% created by:           LaPrint version 3.16 (13.9.2004)
% created on:           31-Aug-2015 16:25:44
% eps bounding box:     16 cm x 12 cm
% comment:              
%
%\begin{psfrags}%
%\psfragscanon%
%
% text strings:
\psfrag{s06}[b][b]{\fontsize{8.5}{12.75}\fontseries{m}\mathversion{normal}\fontshape{n}\selectfont \color[rgb]{0,0,0}\setlength{\tabcolsep}{0pt}\begin{tabular}{c}$\rs(\ttau)$ [bits/sec/Hz]\end{tabular}}%
\psfrag{s07}[t][t]{\fontsize{8.5}{12.75}\fontseries{m}\mathversion{normal}\fontshape{n}\selectfont \color[rgb]{0,0,0}\setlength{\tabcolsep}{0pt}\begin{tabular}{c}$\gamma$ [dB]\end{tabular}}%
\psfrag{s11}[][]{\fontsize{10}{15}\fontseries{m}\mathversion{normal}\fontshape{n}\selectfont \color[rgb]{0,0,0}\setlength{\tabcolsep}{0pt}\begin{tabular}{c} \end{tabular}}%
\psfrag{s12}[][]{\fontsize{10}{15}\fontseries{m}\mathversion{normal}\fontshape{n}\selectfont \color[rgb]{0,0,0}\setlength{\tabcolsep}{0pt}\begin{tabular}{c} \end{tabular}}%
\psfrag{s13}[l][l]{\fontsize{8.5}{12.75}\fontseries{m}\mathversion{normal}\fontshape{n}\selectfont \color[rgb]{0,0,0}EM, $\opc = 0.1$}%
\psfrag{s14}[l][l]{\fontsize{8.5}{12.75}\fontseries{m}\mathversion{normal}\fontshape{n}\selectfont \color[rgb]{0,0,0}IM}%
\psfrag{s15}[l][l]{\fontsize{8.5}{12.75}\fontseries{m}\mathversion{normal}\fontshape{n}\selectfont \color[rgb]{0,0,0}EM, $\opc = 0.01$}%
\psfrag{s16}[l][l]{\fontsize{8.5}{12.75}\fontseries{m}\mathversion{normal}\fontshape{n}\selectfont \color[rgb]{0,0,0}EM, $\opc = 0.1$}%
\psfrag{s17}[b][b]{\fontsize{8.5}{12.75}\fontseries{m}\mathversion{normal}\fontshape{n}\selectfont \color[rgb]{0,0,0}\setlength{\tabcolsep}{0pt}\begin{tabular}{c}$\rs(\ttau)$\end{tabular}}%
\psfrag{s18}[t][t]{\fontsize{8.5}{12.75}\fontseries{m}\mathversion{normal}\fontshape{n}\selectfont \color[rgb]{0,0,0}\setlength{\tabcolsep}{0pt}\begin{tabular}{c}$\gamma$\end{tabular}}%
\psfrag{s19}[b][b]{\fontsize{8.5}{12.75}\fontseries{m}\mathversion{normal}\fontshape{n}\selectfont \color[rgb]{0,0,0}\setlength{\tabcolsep}{0pt}\begin{tabular}{c}Zoom\end{tabular}}%
%
% axes font properties:
\fontsize{8.5}{12.75}\fontseries{m}\mathversion{normal}%
\fontshape{n}\selectfont%
%
% xticklabels:
\psfrag{x01}[t][t]{-1.2}%
\psfrag{x02}[t][t]{-1}%
\psfrag{x03}[t][t]{-0.8}%
\psfrag{x04}[t][t]{-0.6}%
\psfrag{x05}[t][t]{-20}%
\psfrag{x06}[t][t]{-15}%
\psfrag{x07}[t][t]{-10}%
\psfrag{x08}[t][t]{-5}%
\psfrag{x09}[t][t]{0}%
\psfrag{x10}[t][t]{5}%
\psfrag{x11}[t][t]{10}%
%
% yticklabels:
\psfrag{v01}[r][r]{3.4}%
\psfrag{v02}[r][r]{3.5}%
\psfrag{v03}[r][r]{3.6}%
\psfrag{v04}[r][r]{3.7}%
\psfrag{v05}[r][r]{3.8}%
\psfrag{v06}[r][r]{0}%
\psfrag{v07}[r][r]{1}%
\psfrag{v08}[r][r]{2}%
\psfrag{v09}[r][r]{3}%
\psfrag{v10}[r][r]{4}%
\psfrag{v11}[r][r]{5}%
\psfrag{v12}[r][r]{6}%
\psfrag{v13}[r][r]{7}%
\psfrag{v14}[r][r]{8}%
\psfrag{v15}[r][r]{9}%
\psfrag{v16}[r][r]{10}%
%
% Figure:
%\resizebox{8cm}{!}{\includegraphics{fig_opt_thr_vs_SNR_AWGN.eps}}%
%\end{psfrags}%
%
% End fig_opt_thr_vs_SNR_AWGN.tex

\centering
\begin{tikzpicture}[scale=1]
\node[anchor=south west,inner sep=0] (image) at (0,0)
{
	\includegraphics[width= \figscale]{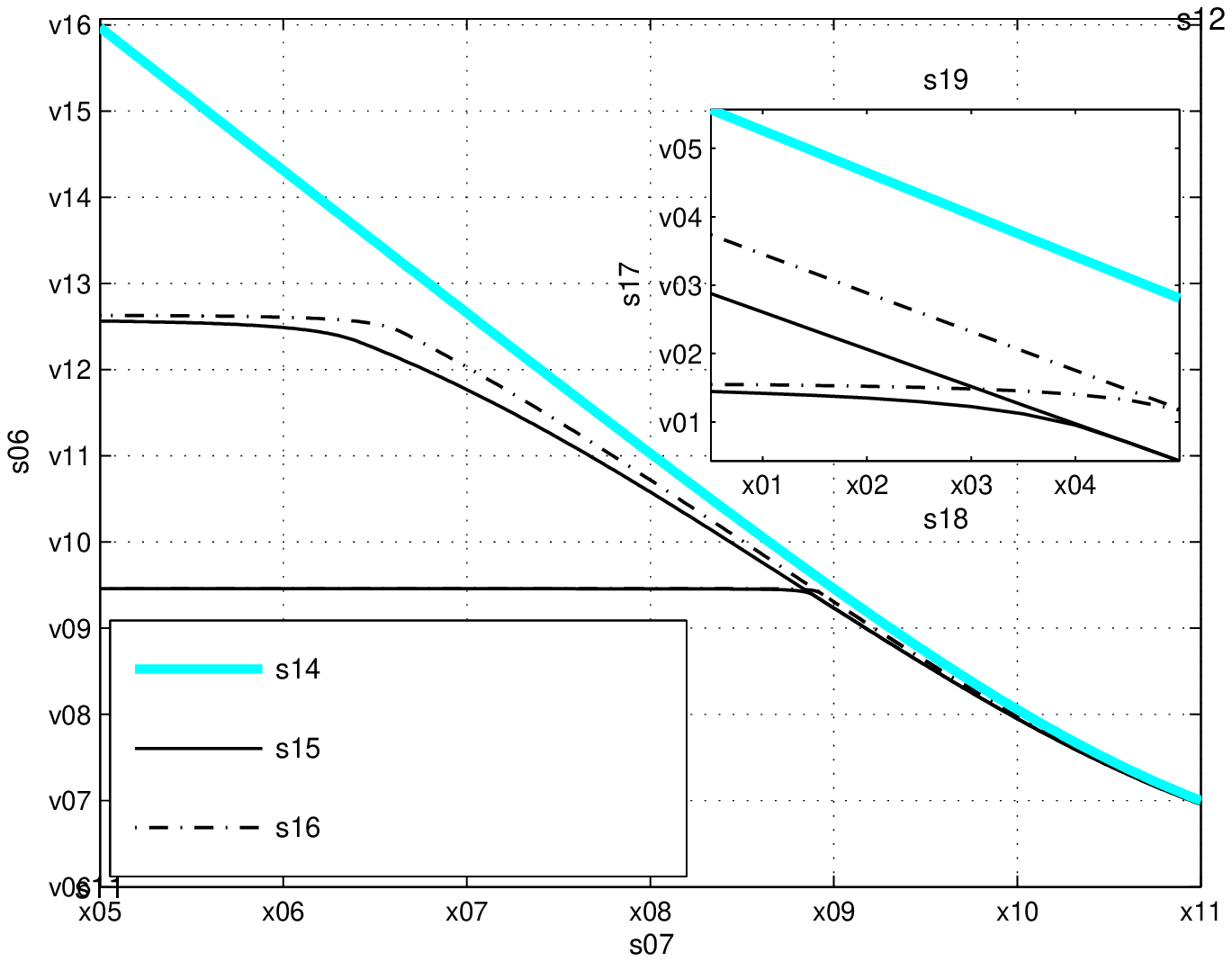}
};
\begin{scope}[x={(image.south east)},y={(image.north west)}]

\draw (0.225,0.65) arc(-160:160:0.01 and 0.03);
\node[draw, fill=gray!10, font=\scriptsize] (text1) at (0.225,0.73) {$\pc = \SI{00}{dBm}$};
\draw (0.225,0.375) arc(-160:160:0.01 and 0.03);
\node[draw, fill=gray!10, font=\scriptsize] (text2) at (0.225,0.455) {$\pc = \SI{-10}{dBm}$};
\draw (0.63,0.66) arc(-150:150:0.016 and 0.048);
\draw[black, <-] (text1.east) -- (0.64,0.73);
\draw (0.63,0.58) arc(-160:160:0.01 and 0.03);
\draw[black, <-] (text2.east) -- (0.635,0.565);

%\draw[help lines,xstep=.1,ystep=.1] (0,0) grid (1,1);
%\foreach \x in {0,1,...,9} { \node [anchor=north] at (\x/10,0) {0.\x}; }
%\foreach \y in {0,1,...,9} { \node [anchor=east] at (0,\y/10) {0.\y}; }
\end{scope}
\end{tikzpicture}
\caption{Optimum throughput $(\rs(\ttau))$ versus the ratio of received control-based power to noise $(\gamma)$ with $\opc \in \{0.01, 0.1\}$ and $\pc \in \{-10, 0\}$ \SI{}{dBm}.}
\label{fig:optT_snr}
\vspace{-7mm}
\end{figure}

\figurename~\ref{fig:ETT} analyzes performance of US in terms of estimation-throughput tradeoff, cf. Theorem \ref{th:th1}, corresponding to the Ideal Model (IM) and the Estimation Model (EM). %Clearly, by relaxing the $\opc$, an improvement in performance it terms of $\trs$ is observed. 
It is indicated that the estimation-throughput tradeoff yields a suitable estimation time $\ttau$ that results in an optimum throughput $\trs(\ttau)$. Hereafter, for the analysis, we consider the theoretical expressions and choose to operate at suitable estimation time. To procure further insights, the variation of $\trs(\ttau)$ with $\gamma$ for different choices of $\pc$ and $\opc$ are considered in \figurename~\ref{fig:optT_snr}. It is observed that $\trs(\ttau)$ gets saturated below a certain $\gamma$, thereby limiting the performance of the US. Particularly for $\pc = \SI{-10}{dBm}$, a severe performance loss indicated by the margin between the IM and the EM is witnessed by the US for $\gamma \le \SI{-2}{dB}$. 

\section{Conclusion} \label{sec:conc}
%%%%%%%%%%%%%%%%%%%%%%%%%%%%%%%%%%%%%%%%%%%%%%%%%%%%%%%%%%%%%%%%%%%%%%%%%%%%%%%%%%%%%%%%%
In this letter, we studied the performance of the USs from a deployment perspective. In this view, a novel model that incorporates channel estimation has been proposed. To capture the impact of imperfect channel knowledge, an outage constraint that forbids performance degradation in terms of interference power received at the primary receiver has been employed. With the inclusion of a transmit power constraint, an operating regime that determines performance limits for the US has been established. Further, a power control mechanism subject to the outage and transmit power constraints has been proposed. Finally, the estimation-throughput tradeoff has been investigated to determine the achievable secondary throughput for the US. In future work, we plan to extend the proposed analysis to include the effect of channel fading in order to characterize the long-term performance of the USs. 
%%%%%%%%%%%%%%%%%%%%%%%%%%%%%%%%%%%%%%%%%%%%%%%%%%%%%%%%%%%%%%%%%%%%%%%%%%%%%%%%%%%%%%%%%
%\appendix[If required]
%%%%%%%%%%%%%%%%%%%%%%%%%%%%%%%%%%%%%%%%%%%%%%%%%%%%%%%%%%%%%%%%%%%%%%%%%%%%%%%%%%%%%%%%%

%%%%%%%%%%%%%%%%%%%%%%%%%%%%%%%%%%%%%%%%%%%%%%%%%%%%%%%%%%%%%%%%%%%%%%%%%%%%%%%%%%%%%%%%%
% References
%%%%%%%%%%%%%%%%%%%%%%%%%%%%%%%%%%%%%%%%%%%%%%%%%%%%%%%%%%%%%%%%%%%%%%%%%%%%%%%%%%%%%%%%%
\bibliographystyle{IEEEtran}
\bibliography{IEEEabrv,refs}

% that's all from my side
\end{document}